\documentclass[twocolumn,showpacs,preprintnumbers,amsmath,amssymb,superscriptaddress,prl]{revtex4}

\usepackage{graphicx}
\usepackage{dcolumn}
\usepackage{bm}

\begin{document}


\title{Observation of Hybrid Soliton Vortex-Ring Structures in Bose-Einstein Condensates}

\author{Naomi S. Ginsberg}
 \email{ginsber@fas.harvard.edu}
\affiliation{ Department of Physics and Division of Engineering
and Applied Sciences, Harvard University, Cambridge, Massachusetts
02138 }

\author{Joachim Brand}
\affiliation{ Max Planck Institute for the Physics of Complex
Systems, 01187 Dresden, Germany }

\author{Lene Vestergaard Hau}
\affiliation{ Department of Physics and Division of Engineering
and Applied Sciences, Harvard University, Cambridge, Massachusetts
02138 }

\date{\today}

\begin{abstract}
We present the experimental discovery of compound structures
comprising solitons and vortex rings in Bose-Einstein condensates
(BECs). We examine both their creation via soliton-vortex
collisions and their subsequent development, which is largely
governed by the dynamics of interacting vortex rings. A
theoretical model in three-dimensional (3D) cylindrical symmetry
is also presented.

\end{abstract}

\pacs{03.75.Lm, 47.37.+q, 47.32.Cc, 42.50.Gy}

\maketitle

Quantized vortices have long been considered a signature of
superfluidity in Helium II \cite{Donnelly}.  The recent emergence
of Bose-condensed alkali gases provides a new and different arena
for studying quantum fluids, their excitations, and interactions
of the latter. In addition to observing vortices alone
\cite{ExpVortex,VortexSpinning1,VortexSpinning2}, experimentalists
have reported observations of grey solitons
\cite{ExpSoliton1,ExpSoliton2} and their decay into vortex rings
(VR) \cite{JILARing,SciSol} via the Kadomtsev-Petviashvili, or
``snake", instability \cite{SnakeCombo}.

In this Letter, we present the observation of a new class of
topological excitations in BECs: hybrid structures originating
from a combination of solitons and VRs. By `soliton', we mean a
non-dispersive nonlinear wave whose phase profile and velocity are
determined by its amplitude. Solitonic wavefronts extend on 2D
sheets whereas vortex structures have a depleted core and phase
singularity along a 1D filament, which may loop to form a VR. We
detail how these nonlinear excitations initially collide and how
the subsequent compound structure dynamics diverge from those of
the superposed evolution of individual excitations. To date, the
interplay between solitons and vortices has only been explored
theoretically \cite{BrandCombo,Komineas3,Berloff}.

In previous work \cite{SciSol}, we studied grey solitons and
vortices generated from the quantum analog of shock waves by
creating a single, narrow density defect in a BEC.  We presently
optimize the likelihood of collisions between these excitations by
creating two symmetrically placed defects via a mechanism similar
to that used in \cite{SciSol}.

\begin{figure}
\includegraphics{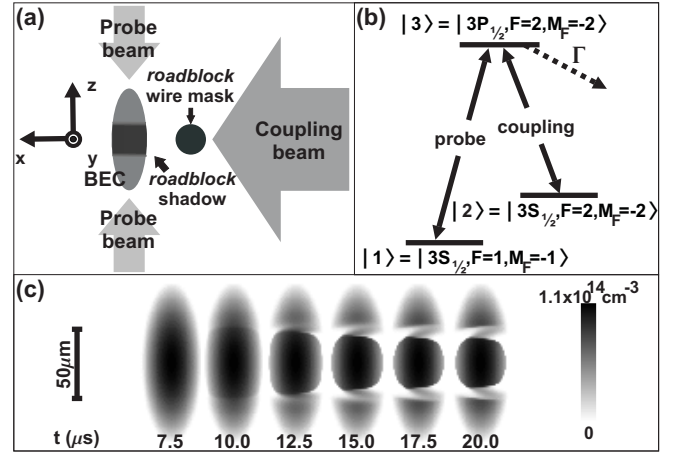}
\caption{\label{fig:1} Defect creation. (a) Double light-roadblock
set-up. (b) Energy level diagram; $M_F$ refers to the projection
of the atoms' angular momentum along the magnetic field
quantization axis, {\it z}, of the 4-Dee electromagnetic trap
\cite{BECproduction}. (c) Density plots of $|1\rangle$ atoms from
2D simulations of defect creation using a 60 $\mu$m wire mask;
initially, the BEC has 4.0$\times$$10^6$ atoms. Light pulse
intensities and durations are as in the experiment.}
\end{figure}

A thin wire blocks the `coupling' beam from illuminating the
center of a magnetically trapped BEC (Fig.\,1(a)), to create a
symmetric {\it double light-roadblock}. We then inject two
counter-propagating Gaussian-shaped `probe' pulses along the
symmetry axis of the trap. The pulses compress spatially by
several orders of magnitude via the slow-light technique
\cite{Nature1}, and ultimately fit entirely inside the BEC. The
probe beams are resonant with the atoms' internal
$|1\rangle$-$|3\rangle$ transition while the coupling beam is
resonant with the $|2\rangle$-$|3\rangle$ transition (Fig.\,1(b)).

When the pulses arrive at the wire's shadow, their propagation is
effectively arrested and they are further compressed, as their
speed depends on the coupling beam intensity that rapidly falls to
zero at the borders of this region \cite{SciSol}. Subsequently,
atoms at the halted pulses' locations are ejected, having first
been transferred from initial state, $|1\rangle$, to $|2\rangle$
(untrapped) via a slow-light induced dark state. This results in
two deep, narrow density defects in $|1\rangle$, symmetric about
the BEC's center (Fig.\,1(c)).

For the current experiment, we condense 3.6$\times$$10^6$ sodium
atoms in $|1\rangle$ and employ magnetic trap frequencies
$\omega_z$=2$\pi$$\times$21 Hz and
$\omega_x$=$\omega_y$=3.0$\omega_z$. We work at temperatures well
below the transition temperature to BEC so there is virtually no
non-condensed component. The coupling (peak probe) Rabi frequency
is 2$\pi$$\times$15.3 (3.8) MHz, and duration (probe 1/e
half-width) is 12 (2.5) $\mu$s.  We use a 35 $\mu$m wire mask.

After creating the defects, we vary the duration of condensate
evolution in the magnetic trap, $\Delta t_{trap}$, and then
release the condensate and let it expand for $\Delta t_{exp}$=19.9
ms before using a slice technique \cite{slice,SciSol} to image the
central 20-30 $\mu$m thick slab of the expanded cloud in the {\it
x-z} plane. Figure 2(a) shows the `control' experiment in which a
single probe pulse is injected into the BEC, while Fig.\,2(b)
depicts the double-roadblock experiment in which interactions
between excitations are evident. In both cases, the data are
highly reproducible.

The early dynamics that ensue from the creation of a single defect
are detailed in Fig.\,3 of \cite{SciSol}. The defect splits into
two density dips propagating at the sound speed in the $+$$z$ and
$-$$z$ directions, while the back edge of each steepens due to the
density dependence of the sound speed. In a classical fluid, this
would lead to shock wave formation i.e.\,the back edges would
develop infinite slope. In a superfluid, significant density
changes over distances shorter than the healing length cannot
occur \cite{PandS}.

Consequently, as the density dips travel, they shed grey solitons
each time their back edges become too steep \cite{SciSol}. As
observed in Fig.\,2(a), the solitons, although stable in 1D, kink
and decay into vortex structures via the snake instability, seeded
predominantly by the transverse density variation of the
condensate depicted in Fig.\,1(c).

\begin{figure*}
\includegraphics{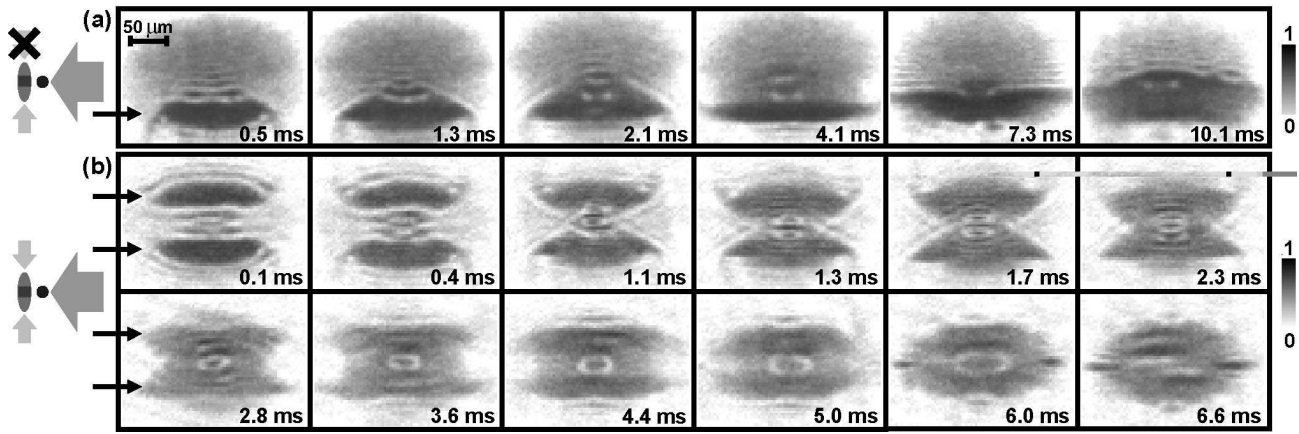}
\caption{\label{fig:2}Experimentally observed resonant
transmission images of BECs (transmitted laser intensity
normalized to incident) after they are illuminated with coupling
and probe pulse(s). In (a) ((b)), one (two) probe pulse(s) is
(are) hitting the cloud, there are initially 4.0 (3.6)$\times
10^6$ condensed atoms, and the imaged slice is 30 (23) $\mu$m
thick. Times cited refer to evolution time in trap, $\Delta
t_{trap}$, after defect creation; $\Delta t_{exp}$$=$19.9 ms.
Arrows indicate initial defect planes.}
\end{figure*}

We now describe the experimental data in Fig. 2, comparing and
contrasting the double-defect images with the control experiment,
before considering a corresponding theoretical simulation.
Focusing mostly on structures between the two initial defects, we
see at $\Delta t_{trap}$=0.1 ms in Fig.\,2(b) that solitons (pale
stripes) have already shed from density dips originally emanating
from the defects. High-density bands on the top and bottom result
from the rush of fluid into the defect locations. The first
solitons shed are the deepest, hence slowest moving, and therefore
bound the regions of high density. In the center, between these
first solitons, secondly shed ones have kinked so that they
overlap on the sides but not in the middle, forming a checkerboard
pattern. This is clearly seen at $\Delta t_{trap}$=0.4 ms where
the images also show VR cores in the `four corners'. Figure 2(b)
at 0.1 and 0.4 ms and Fig.\,2(a) at 0.5 ms are comparable. The
single-defect image shows additional shallower solitons in the
wake of the travelling density dip, that are initially hidden in
the double-defect version by the mirror image of higher-contrast
structures. At $\Delta t_{trap}$=1.1 ms (Fig.\,2(b)), soliton
curvature has increased even more. Like at 1.3 ms of Fig.\,2(a),
the deepest solitons form double-cusped shapes (`W' on the top,`M'
on the bottom) that span the cloud from side to side. The second
pair of solitons propagate toward one another further, closing off
a region of fluid in the center.

We have explored the 3D structure of the excitations by varying
the $y$-position of the imaged slice to find that the system is
highly cylindrically symmetric. Thus, the white loop in the center
at 1.1 ms and the pairs of flattened loops seen in the shots
corresponding to $\Delta t_{trap}$=1.3 and 1.7 ms represent
roughly ellipsoidal {\it shells} of low density, a feature clearly
absent in the single-defect case. Comparing the latter two
time-points with $\Delta t_{trap}$=2.1 ms in Fig.\,2(a), we
conclude that the centers of the deepest solitons have decayed
into VRs, forming the equatorial components of their corresponding
closed low density shells in Fig.\,2(b). The upper and lower
`hemispheres' of the shells all derive from solitons formed after
the first (deepest) ones that produced VRs.

At $\Delta t_{trap}$=2.3 ms in Fig.\,2(b), the closed structures
appear to interlace and subsequently form a single low density
shell at the center of the condensate (2.8-6.0 ms). This provides
very strong evidence that the double-defect dynamics are not
merely a superposition of mirrored single-defect dynamics. At
$\Delta t_{trap}$=4.1 ms, Fig.\,2(a) displays a soliton fragment
in the center with a VR below.  At 4.4 ms in Fig.\,2(b), the low
density structure is {\it closed} and could not be formed from the
superposition of elements from the corresponding time point in
Fig.\,2(a). Despite being larger at $\Delta t_{trap}$=5 and 6 ms
(Fig.\,2(b)) than in previous frames, this compound structure is
temporarily stable. At 6.6 ms, we observe that what was a closed
shell has sheared and come apart at the sides. The resulting
structures break cylindrical symmetry, highlighting their
sensitivity to torque from the laser beams, evident in Fig.\,1(c).
Furthering the case for interactions rather than superposition,
the last two frames of Fig.\,2(a) have a long-lived VR in the
center, in stark contrast with Fig.\,2(b) at 6.6 ms, which has
segments that could not be constructed from VRs alone.

We measure the soliton and shell half-widths to
range from 3 to 6 $\mu$m, comparable to a calculated healing length of 4 
$\mu$m \cite{CastinDum}. We also note the collective modes in the
BEC produced by the initial rush of fluid into the defects'
locations. The high density bands in both Figs.\,2(a) and 2(b)
bend, contract, and expand. We find these modes are highly
dependent on the initial defect shape.

To complement our observations, we have performed simulations
based upon a generalized Gross-Pitaevskii (GP) description
\cite{processing} for three phases of BEC evolution as follows: We
first simulate defect creation via slow-light propagation on a 2D
spatial grid, as in Fig.\,1(c). The atoms in $|2\rangle$ are
quickly ejected from the trap and we map a symmetrized version of
the resulting defect in the $|1\rangle$ condensate onto a 3D
cylindrically symmetric grid \cite{symm}. We then calculate
in-trap evolution of the $|1\rangle$ atoms for a duration $\Delta
t_{trap}$ and finally continue to follow the BEC evolution after
the trapping potential is turned off.

\begin{figure}
\includegraphics{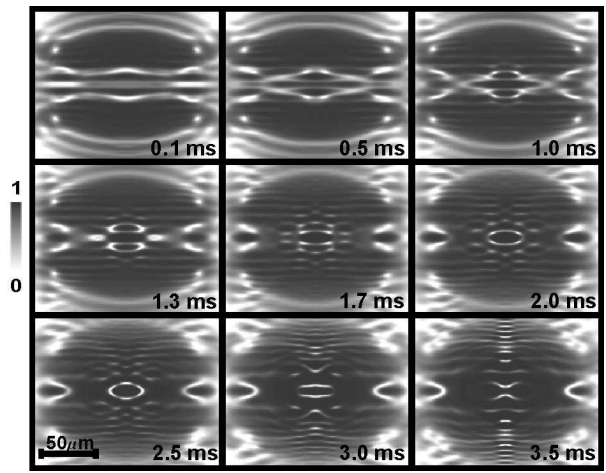}
\caption{\label{fig:3} Simulation for $\Delta t_{exp}$=14 ms
($\Delta t_{trap}$ on plots) with parameters the same as in
Fig.\,1(c). To compare with experiment, we plot $e^{-D_y(x,z)}$,
where $D_y(x,z)$ is the optical density along $y$ over the central
22 $\mu$m slab of the BEC.}
\end{figure}

The results of simulations for $\Delta t_{exp}$=14 ms with varying
$\Delta t_{trap}$ are shown in Fig. 3. (After 14 ms of cloud
expansion, all topological dynamics are frozen out and further
evolution simply leads to spatial magnification of the developed
features; the BEC's aspect ratio differs from that at 19.9 ms by
$\sim$15\%.) We see the same prominent features in the simulation
as in the experiment, allowing us to identify the nature of the
low density shells. Phase information confirms the existence of
solitons and identifies the creation and annihilation of VRs. As
we detail below, the simulations indicate that the expansion
`restabilizes' solitons that would have been subjected to the
snake instability, if kept in the trap. Experimental parameters
have been chosen to cover a regime in which VRs are produced but
parent solitons are not destroyed.

\begin{figure*}
\includegraphics{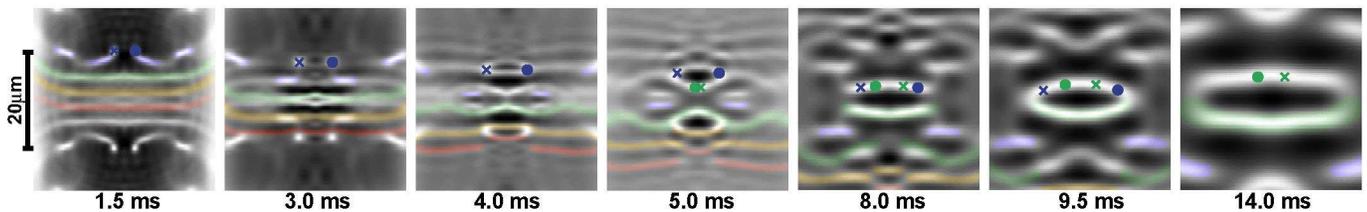}
\caption{\label{fig:4} Simulation of soliton VR dynamics during
expansion after $\Delta t_{trap}$=3.0 ms. Times cited are $\Delta
t_{exp}$. Only solitons propagating in $-$$z$ direction (downward)
from the $+$$z$ (top) defect are shaded (blue through red, in
order of creation), to assist in tracking the movement of each
from frame to frame. Dots (crosses) indicate vortex cores for
$z$$>$0 {\it only} with positive (negative) circulation with
respect to the plane of the page. The vortex colors correspond to
parent soliton colors, although the green cores are generated by
the mirror image of the correspondingly shaded soliton. Density
ranges from 0 to \{7.5, 4.5, 4.1, 3.6, 1.3, 0.92,
0.42\}$\times$$10^{13}cm^{-3}$ for each of the times presented,
respectively. Simulation parameters are the same as in
Fig.\,1(c).}
\end{figure*}

The simulation follows that of the experiment up to $\Delta
t_{trap}$$\sim$1.3 ms, after which point evolution is slightly
accelerated.  Other discrepancies include reduced soliton
curvature and less dramatic collective mode breathing than in the
experiment. This could be a result of the different initial defect
separation used in the theory to generate a sequence most
comparable to the data. The simulation's cylindrical symmetry is
an added approximation, but it generally describes the
observations well.

Despite these differences, the simulations provide us with the
history of the experimentally observed structures that must be
expanded to be optically resolved. Calculations show that if two
solitons are incident on one another, the time that it takes them
to reach each other and their degree of kinking at that point have
different density dependencies. During expansion but before
interaction energy is completely converted to kinetic, soliton
translational velocity and the rate of soliton kinking both
decrease as BEC density drops, but at different rates. These
different density dependencies do not only imply that expansion
evolution is highly important and different than in-trap evolution
\cite{singleDefCaveat}; they also lead to the notion that
expansion sequences succeeding two different in-trap durations are
not mere duplicates offset from each other in time. We see
variations in such sequences, typically manifested by the
occurrence of VR creation and annihilation at slightly different
points in the evolution.

As an example, we present the simulated release dynamics for
$\Delta t_{trap}$=3.0 ms in Fig.\,4. As in the experiment, the
first soliton (blue) shed from each of the waves travelling toward
the BEC's center is the deepest, slowest, and least stable,
leading to decay into VRs at $\Delta t_{exp}$=1.5 ms. Solitons
created after the first ones (green through red, in order of
creation from top defect) propagate across the center line of the
BEC ($\Delta t_{exp}$=1.5 and 3.0 ms); those with opposite
propagation directions sometimes overlap. As the third and fourth
generated solitons reach the newly formed VRs on the side opposite
whence they originated, we observe pairs of low density shells
($\Delta t_{exp}$=4.0 ms). By examining the phase, we confirm, as
discussed for images 1.3 and 1.7 ms of Fig.\,2(b), that these
shells are made from VRs and transiently passing soliton fronts.

The second deepest (secondly shed, green) solitons cross in the
center of the BEC as their midpoints bend toward one another (also
$\Delta t_{exp}$=4.0 ms). As these fronts reach the VRs at 5.0 ms
(again forming a pair of low density shells), the central portions
of the solitons curve further, producing within themselves two
smaller VRs with opposite circulation from the larger ones. These
truly hybrid low density structures, consisting of parent soliton
segments and VRs, `dislocate' from the rest of the soliton fronts.
They each comprise two concentric VRs embedded in a roughly
hemispherical solitonic shell. The smaller VRs are evident from
the wavefunction's phase even though they are not distinguishable
in the density plots. The general motion of the hybrid segments is
largely determined by the velocity fields from the vortices
present. Like an umbrella turning inside-out, the curvature of
these hybrid structures reverses through $\Delta t_{exp}$=8.0 ms,
as the inner rings propel themselves away from one another and the
outer rings propel toward the center line. This occurs in the
experiment for $\Delta t_{trap}$=2.3 ms and in Fig.\,3 at 1.7 ms.
The VRs behave as the `phantom' propellers in the compound
structures.

As the outer VRs propagate toward each other, a closed shell forms
from the two hybrid segments ($\Delta t_{exp}$=9.5 ms). It is an
ellipsoidal solitonic shell incorporating four VRs of alternating
circulation that encircle the structure like lines of constant
latitude on a globe. The outer VRs eventually annihilate
(quantized circulation at the VR cores disappears). Subsequently,
the two segments that formed the central shell lose their
curvature, and the shell breaks up along {\it z}=0. This is seen
at $\Delta t_{exp}$=14 ms (a close-up of the BEC at $\Delta
t_{trap}$=3.0 ms in Fig.\,3) and should be compared to the
experimentally observed shearing at 6.6 ms in Fig.\,2(b).

In light of these observations it is natural to ask whether low
density shells can exist as stationary states in a BEC. We have
shown that the GP equation exhibits stationary solutions in
spherical symmetry consisting of an inner sphere of almost
constant density surrounded by low density shells close to
Bessel-function type spherical standing waves $\sin(k r)/r$ known
from the linear Schr\"odinger equation. This has been
independently treated by others \cite{shells}. The infinite space
solutions have an infinite number of particles and concentric
shells. In a trap, similar solutions with finite particle number
and number of shells are possible; details will be given
elsewhere. Unlike the observed structures discussed above, the
spherical stationary states show no vortex core structures.
However, the observed hybrid shells might play a role in the decay
of the stationary solutions.

In brief, we have presented the experimental observation and
theoretical confirmation of low density shells in a fully 3D BEC,
consisting of complex hybrid soliton vortex-ring structures. VR
propulsion, attraction, and annihilation heavily influence the
structures' dynamics. The BEC's decreasing density, after a
confining trap is switched off, contributes critically to the
dynamics, as the resulting regime is on the cusp of stability of
solitons typically subject to the snake instability in 3D.

\begin{acknowledgments}
The authors thank Z. Dutton and S. Komineas for fruitful
discussion and insights. This work was supported by AFOSR, NSF,
NASA, and the ARO-MURI Program.
\end{acknowledgments}


\begin{thebibliography}{21}
\expandafter\ifx\csname
natexlab\endcsname\relax\def\natexlab#1{#1}\fi
\expandafter\ifx\csname bibnamefont\endcsname\relax
  \def\bibnamefont#1{#1}\fi
\expandafter\ifx\csname bibfnamefont\endcsname\relax
  \def\bibfnamefont#1{#1}\fi
\expandafter\ifx\csname citenamefont\endcsname\relax
  \def\citenamefont#1{#1}\fi
\expandafter\ifx\csname url\endcsname\relax
  \def\url#1{\texttt{#1}}\fi
\expandafter\ifx\csname
urlprefix\endcsname\relax\def\urlprefix{URL }\fi
\providecommand{\bibinfo}[2]{#2}
\providecommand{\eprint}[2][]{\url{#2}}

\bibitem[{\citenamefont{Donnelly}(1991)}]{Donnelly}
\bibinfo{author}{\bibfnamefont{R.~J.} \bibnamefont{Donnelly}},
  \emph{\bibinfo{title}{Quantized Vortices in Helium II}}
  (\bibinfo{publisher}{Cambridge Univ. Press}, \bibinfo{address}{Cambridge,
  UK}, \bibinfo{year}{1991}).

\bibitem[{\citenamefont{Matthews et~al.}(1999)}]{ExpVortex}
\bibinfo{author}{\bibfnamefont{M.~R.} \bibnamefont{Matthews}}
  \bibnamefont{et~al.}, \bibinfo{journal}{Phys. Rev. Lett.}
  \textbf{\bibinfo{volume}{83}}, \bibinfo{pages}{2498} (\bibinfo{year}{1999}).

\bibitem[{\citenamefont{Madison et~al.}(2000)\citenamefont{Madison, Chevy,
  Wohlleben, and Dalibard}}]{VortexSpinning1}
\bibinfo{author}{\bibfnamefont{K.~W.} \bibnamefont{Madison}},
  \bibinfo{author}{\bibfnamefont{F.}~\bibnamefont{Chevy}},
  \bibinfo{author}{\bibfnamefont{W.}~\bibnamefont{Wohlleben}},
  \bibnamefont{and} \bibinfo{author}{\bibfnamefont{J.}~\bibnamefont{Dalibard}},
  \bibinfo{journal}{Phys. Rev. Lett.} \textbf{\bibinfo{volume}{84}},
  \bibinfo{pages}{806} (\bibinfo{year}{2000}).

\bibitem[{\citenamefont{Abo-Shaeer et~al.}(2001)\citenamefont{Abo-Shaeer,
  Raman, Vogels, and Ketterle}}]{VortexSpinning2}
\bibinfo{author}{\bibfnamefont{J.~R.} \bibnamefont{Abo-Shaeer}},
  \bibinfo{author}{\bibfnamefont{C.}~\bibnamefont{Raman}},
  \bibinfo{author}{\bibfnamefont{J.~M.} \bibnamefont{Vogels}},
  \bibnamefont{and} \bibinfo{author}{\bibfnamefont{W.}~\bibnamefont{Ketterle}},
  \bibinfo{journal}{Science} \textbf{\bibinfo{volume}{292}},
  \bibinfo{pages}{476} (\bibinfo{year}{2001}).

\bibitem[{\citenamefont{Burger et~al.}(1999)}]{ExpSoliton1}
\bibinfo{author}{\bibfnamefont{S.}~\bibnamefont{Burger}} \bibnamefont{et~al.},
  \bibinfo{journal}{Phys. Rev. Lett.} \textbf{\bibinfo{volume}{83}},
  \bibinfo{pages}{5198} (\bibinfo{year}{1999}).

\bibitem[{\citenamefont{Denschlag et~al.}(2000)}]{ExpSoliton2}
\bibinfo{author}{\bibfnamefont{J.}~\bibnamefont{Denschlag}}
  \bibnamefont{et~al.}, \bibinfo{journal}{Science}
  \textbf{\bibinfo{volume}{287}}, \bibinfo{pages}{97} (\bibinfo{year}{2000}).

\bibitem[{\citenamefont{Anderson et~al.}(2001)}]{JILARing}
\bibinfo{author}{\bibfnamefont{B.~P.} \bibnamefont{Anderson}}
  \bibnamefont{et~al.}, \bibinfo{journal}{Phys. Rev. Lett.}
  \textbf{\bibinfo{volume}{86}}, \bibinfo{pages}{2926} (\bibinfo{year}{2001}).

\bibitem[{\citenamefont{Dutton et~al.}(2001)\citenamefont{Dutton, Budde, Slowe,
  and Hau}}]{SciSol}
\bibinfo{author}{\bibfnamefont{Z.}~\bibnamefont{Dutton}},
  \bibinfo{author}{\bibfnamefont{M.}~\bibnamefont{Budde}},
  \bibinfo{author}{\bibfnamefont{C.}~\bibnamefont{Slowe}}, \bibnamefont{and}
  \bibinfo{author}{\bibfnamefont{L.~V.} \bibnamefont{Hau}},
  \bibinfo{journal}{Science} \textbf{\bibinfo{volume}{293}},
  \bibinfo{pages}{663} (\bibinfo{year}{2001}).

\bibitem[{Sna()}]{SnakeCombo}
\bibinfo{note}{B. B. Kadomtsev and V. I. Petviashvili, Sov. Phys. Dokl. {\bf
  15}, 539 (1970); C. A. Jones, S. J. Putterman, P. H. Roberts, J. Phys. A {\bf
  19}, 2991 (1986); C. Josserand and Y. Pomeau, Europhys. Lett. {\bf 30}, 43
  (1995); D. L. Feder et al., Phys. Rev. A {\bf 62}, 053606 (2000); A. Muryshev
  et al., Phys. Rev. Lett. {\bf 89}, 110401 (2002).}

\bibitem[{Bra()}]{BrandCombo}
\bibinfo{note}{J. Brand and W. P. Reinhardt, J. Phys. B {\bf 34}, L113 (2001);
  J. Brand and W. P. Reinhardt, Phys. Rev. A {\bf 65}, 043612 (2002).}

\bibitem[{\citenamefont{Komineas and Papanicolaou}(2003)}]{Komineas3}
\bibinfo{author}{\bibfnamefont{S.}~\bibnamefont{Komineas}} \bibnamefont{and}
  \bibinfo{author}{\bibfnamefont{N.}~\bibnamefont{Papanicolaou}},
  \bibinfo{journal}{Phys. Rev. A} \textbf{\bibinfo{volume}{68}},
  \bibinfo{pages}{043617} (\bibinfo{year}{2003}).

\bibitem[{\citenamefont{Berloff}(2004)}]{Berloff}
\bibinfo{author}{\bibfnamefont{N.~G.}~\bibnamefont{Berloff}},
  \bibinfo{journal}{Phys. Rev. A} \textbf{\bibinfo{volume}{69}},
  \bibinfo{pages}{053601} (\bibinfo{year}{2004}).

\bibitem[{\citenamefont{Hau et~al.}(1998)}]{BECproduction}
\bibinfo{author}{\bibfnamefont{L.~V.} \bibnamefont{Hau}} \bibnamefont{et~al.},
  \bibinfo{journal}{Phys. Rev. A} \textbf{\bibinfo{volume}{58}},
  \bibinfo{pages}{R54} (\bibinfo{year}{1998}).

\bibitem[{\citenamefont{Hau et~al.}(1999)\citenamefont{Hau, Harris, Dutton, and
  Behroozi}}]{Nature1}
\bibinfo{author}{\bibfnamefont{L.~V.} \bibnamefont{Hau}},
  \bibinfo{author}{\bibfnamefont{S.~E.} \bibnamefont{Harris}},
  \bibinfo{author}{\bibfnamefont{Z.}~\bibnamefont{Dutton}}, \bibnamefont{and}
  \bibinfo{author}{\bibfnamefont{C.~H.} \bibnamefont{Behroozi}},
  \bibinfo{journal}{Nature} \textbf{\bibinfo{volume}{397}},
  \bibinfo{pages}{594} (\bibinfo{year}{1999}).

\bibitem[{\citenamefont{Andrews et~al.}(1997)}]{slice}
\bibinfo{author}{\bibfnamefont{M.}~\bibnamefont{Andrews}} \bibnamefont{et~al.},
  \bibinfo{journal}{Science} \textbf{\bibinfo{volume}{275}},
  \bibinfo{pages}{637} (\bibinfo{year}{1997}).

\bibitem[{\citenamefont{Pethick and Smith}(2002)}]{PandS}
\bibinfo{author}{\bibfnamefont{C.~J.} \bibnamefont{Pethick}} \bibnamefont{and}
  \bibinfo{author}{\bibfnamefont{H.}~\bibnamefont{Smith}},
  \emph{\bibinfo{title}{Bose-Einstein Condensation in Dilute Gases}}
  (\bibinfo{publisher}{Cambridge Univ. Press}, \bibinfo{address}{UK},
  \bibinfo{year}{2002}).

\bibitem[{\citenamefont{Castin and Dum}(1996)}]{CastinDum}
\bibinfo{author}{\bibfnamefont{Y.}~\bibnamefont{Castin}} \bibnamefont{and}
  \bibinfo{author}{\bibfnamefont{R.}~\bibnamefont{Dum}},
  \bibinfo{journal}{Phys. Rev. Lett.} \textbf{\bibinfo{volume}{77}},
  \bibinfo{pages}{5315} (\bibinfo{year}{1996}).

\bibitem[{pro()}]{processing}
\bibinfo{note}{Z. Dutton and L. V. Hau, Phys. Rev. A in press,
  quant-ph/0404018.}

\bibitem[{sym()}]{symm}
\bibinfo{note}{Density in the 2D simulation was normalized and then
  interpolated. The result was averaged with its mirror image about $z$ and
  then scaled by the 3D BEC ground state density for the same number of atoms.}

\bibitem[{sin()}]{singleDefCaveat}
\bibinfo{note}{It is not as critical to distinguish between in-trap and
  expansion evolution in the single-defect case since its geometry does not
  lead to the possibility of soliton collisions.}

\bibitem[{she()}]{shells}
\bibinfo{note}{M. Machholm and H. Smith, private communication; L. D. Carr and
  C. W. Clark, cond-mat/0408460.}

\end{thebibliography}

\end{document}